\documentclass[12pt,preprint]{emulateapj}
\usepackage{natbib}
\shorttitle{Modeling The GRB Host Galaxy Mass Distribution}
\shortauthors{D. Kocevski et al.}

\begin{document}

\title{Modeling The GRB Host Galaxy Mass Distribution: Are GRBs Unbiased Tracers of Star Formation?}

\author{Daniel Kocevski \altaffilmark{1}, Andrew A. West \altaffilmark{2},\altaffilmark{3}, \& Maryam Modjaz  \altaffilmark{2},\altaffilmark{4}} 

\altaffiltext{1}{Kavli Institute for Particle Astrophysics and Cosmology, Stanford University, 2575 Sand Hill Road M/S 29, Menlo Park, Ca 94025 }
\altaffiltext{2}{Astronomy Department, University of California, 601 Campbell Hall, Berkeley, CA 94720 }
\altaffiltext{3}{MIT Kavli Institute for Astrophysics and Space Research, 77 Massachusetts Ave, Cambridge, MA, 02139 }
\altaffiltext{4}{Miller Research Fellow, Astronomy Department, University of California, 601 Campbell Hall, Berkeley, CA 94720 }


\begin{abstract}

  We model the mass distribution of long gamma-ray burst (GRB) host
  galaxies given recent results suggesting that GRBs occur in low
  metallicity environments. By utilizing measurements of the redshift
  evolution of the mass-metallicity (M-Z) relationship for galaxies,
  along with a sharp host metallicity cut-off suggested by Modjaz
  and collaborators, we estimate an upper limit on the stellar mass of
  a galaxy that can efficiently produce a GRB as a function of
  redshift.    By employing consistent abundance indicators, we find that sub-solar metallicity cut-offs effectively limit GRBs to low stellar mass spirals and dwarf galaxies at low redshift.  At higher redshifts, as the average metallicity of galaxies in the Universe falls, the mass range of galaxies capable
  of hosting a GRB broadens, with an upper bound approaching the mass
  of even the largest spiral galaxies.  We compare these predicted limits to the
  growing number of published GRB host masses and find that extremely low
  metallicity cut-offs of 0.1 to 0.5 $Z_{\odot}$ are effectively ruled out by a large number of intermediate mass galaxies at low redshift.  A mass function that includes a smooth decrease in the efficiency of producing GRBs in galaxies of metallicity above 12+log(O/H)$_{\rm KK04}$ 8.7 can, however, accommodate a majority of the measured host galaxy masses.  We find that at $z \sim 1$, the peak in the observed GRB host
  mass distribution is inconsistent with the expected peak in the mass of galaxies harboring most of the star formation.  This suggests that GRBs are metallicity biased tracers of star formation at low and intermediate redshifts, although our model predicts that this bias should disappear at higher redshifts due to the evolving metallicity content of the universe.

\end{abstract}


\section{Introduction}

The success of NASA's Swift spacecraft \citep{Gehrels04} has resulted
in a dramatic increase in the number of X-ray localizations of afterglows associated with long gamma-ray bursts (GRBs).  This
increase has resulted in a wealth of new information regarding the
demographics of GRBs and their host galaxies.  Investigating the
environments in which these events occur has long been an important
path to understanding the nature of GRB progenitors, as different
origin models have traditionally predicted distinct GRB host galaxy populations. The connection between GRBs and the death of massive stars is now
 well-established at low redshift ($z < 0.3$) by the association of GRBs
 with broad lined SN lc events \citep[for a review, see][]{Woosley06}.


Recent observations \citep{Ceron08, Savaglio09} of X-ray localizations
by Swift have bolstered previous results showing that GRB host
galaxies tend to be bluer, fainter, and more irregular than
$M_{\star}$ galaxies at similar redshifts \citep{Fruchter99, Chary02,
  Bloom02, LeFloch03, Tanvir04, Fruchter06, Ceron06}.  They tend to
have higher specific star formation than typical star-forming galaxies
\citep{Chary02, Berger03, Christensen04} and the peak in their redshift
distribution tends to broadly track the peak in the overall cosmic star formation rate of
the universe \citep{Bloom03, Firmani04, Natarajan05, Jakobsson06,
  Kocevski06, Guetta07}.  Only a handful of events have been
associated with grand design spirals and no long duration GRB has been
associated with an early type galaxy.

A growing body of spectroscopic evidence has also shown that these
galaxies tend to be metal poor \citep{Prochaska04, Sollerman05, Fruchter06, Modjaz06, Stanek07, Thoene07, Wiersema07, Margutti07}. Absorption line spectroscopy has revealed that the regions in which GRB afterglows are observed tend to
have the low metallicities that are expected from young stellar
populations \citep{Fynbo03, Savaglio03}.  However, there
are a few exceptions \citep{Fynbo06, Prochaska07, Fynbo08, Chen08}.  The high specific star formation rates along with these low metallicities
are similar to what are seen in star-bursting Lyman break galaxies at high redshift.

There is ample theoretical justification for the {\it a priori}
association of GRBs with short-lived, metal poor progenitors.  The combination of
high angular momentum and high stellar mass at the time of collapse
\citep{Woosley93, MacFadyen99} is crucial for producing the collimated
emission that is required to account for the enormous
isotropic-equivelent energy released by these events.  Low metallicity
progenitors would, in theory, retain more of their mass due to smaller
line-driven stellar winds \citep{Kudritzki00, Vink05}, and hence preserve their angular momentum
\citep{Yoon05, WoosleyHeger06}.

Recently, \citet{Modjaz08} showed that a sharp delineation may exist
between the metallicity at the sites of broad-lined SN Ic that
have been associated with GRBs and SN Ic with no detected gamma-ray
emission.  Using a sample of 12 nearby ($z<0.14$) broad-lined SN Ic without
associated GRBs, they found that the chemical abundance at the sites of
known SN-GRBs (at $z<0.25$) were systematically lower than those
harboring SN without GRBs, with a boundary between the two samples at
an oxygen abundance of roughly 12+log(O/H)$_{\rm KD02} \sim 8.5$ in the
\citet{KD02} scale (see Modjaz et al$.$ 2008 - Figure 5).  This trend is
independent of choice of the metallicity diagnostic they adopt (see their
Figure 6) and the mode of SN survey that found the SN without GRBs.

At the same time, the observed trend that many GRB host galaxies are
less luminous, metal poor, irregular dwarf galaxies is in qualitative
agreement with the observed trend of decreasing metallicity of
galaxies as a function of their stellar mass: the mass-metallicity
(M-Z) relationship.  Although well established at low redshift
\citep{Tremonti04}, the M-Z relationship has only recently been
measured for high redshift galaxies where it has become clear that the
overall normalization of the relationship has decreased throughout the
history of the universe \citep{Savaglio05, Erb06}.

As a consequence of the M-Z relationship, any bias in the metallicity
of the environment that is capable of producing a GRB would likely
place severe restrictions on the type of galaxies that can host such
events.  While earlier studies suggest that the GRB redshift distribution tends to broadly track the overall cosmic star formation rate of the universe, the question remains as to what extent GRB hosts are unbiased tracers of SF in the high redshift universe.  In this paper, we use empirical models based on the
measurements of the redshift evolution of the M-Z relationship to
estimate the upper limit to the stellar mass of a galaxy that can
harbor a GRB, and test the suggestion that GRBs preferentially form in low metallicity
environments.  We detail the prescriptions for our model in $\S 2$ and
expand upon our results in $\S 3$.  We compare our model predictions
to published host mass values in $\S 4$ and discuss the implications
of our results in $\S 5$.

\section{Model Prescriptions} \label{sec:Models}

To investigate how a potential metallicity cut-off effects the
resulting GRB host mass distribution we must first assume an empirical
prescription for the relationship between a galaxy's stellar mass and
its level of chemical enrichment.  Such a correlation was first
observed by Lequeux et al. (1979); a trend between the heavy-element
abundance in H II regions and the stellar mass of irregular and blue
compact galaxies.  More recently, this correlation has been
statistically quantified by \citet{Tremonti04} using of $\sim$ 53,000
galaxies from the Sloan Digital Sky Survey (SDSS).  \citet{Tremonti04}
find a tight correlation between galactic stellar mass and gas-phase
metallicity that spans 3 orders of magnitude in stellar mass and a
factor of 10 in metallicity.  They conclude that the galactic
metallicity abundance rises steeply for stellar masses between
$10^{8.5}$ and $10^{10.5}$ M$_{\odot}$, then flattens for galaxies
above $10^{10.5}$ M$_{\odot}$.


A basic form of this correlation is a natural consequence of the
conversion of gas to stars within star forming galaxies, given a
mass dependent star formation efficiency \citep{Schmidt63, Searle72}.
In the context of these simple ``closed-box'' models, this disparity
in the efficiency between high and low mass galaxies is thought to be
due to variations in galactic surface densities as a function of mass
\citep{Kennicutt98, Martin01, Dalcanton04}. 

It has now become apparent
that the effects of supernovae feedback and the infall of metal-poor
gas \citep{Dalcanton07} must also play important roles in shaping the
observed mass-metallicity relationship.  Galactic winds produced by
SNe work to strip galaxies of metal enriched gas, with low mass
galaxies being more susceptible to such effects due to their shallower
potential wells.  Energy injection from SNe also heats interstellar
gas, delaying the collapse of otherwise cold gas to produce stars.  At
the same time, the infall of metal-poor gas acts to dilute the metal
content of the ISM.  This effect is significant in small galaxies
where the infall rate can exceed the total star formation rate,
causing an net decrease in the metallically of the ISM with time.  The
combined result of these mechanisms is that high mass galaxies process
their primordial gas faster and more efficiently than low mass
galaxies and are more effective at retaining the resulting material
against wind induced mass loss, leading to a positive correlation
between stellar mass and metallicity.

These explanations for the origin of the M-Z relationship suggest
significant evolution of the relationship with redshift.  First,
one would expect the normalization of the relationship to fall as a function of lookback time as metal abundance
becomes less common in all galaxies.  Second, the variations in the
efficiency of star formation as a function of mass should also change
the slope of the M-Z relationship as a function of redshift.  Efforts
to quantify this evolution have been the focus of several recent
observational \citep{Savaglio05, Erb06, Maiolino08} and numerical
\citep{Rossi07, Kobayashi07, Tassis08, Brooks07} investigation.  In
particular, \citet{Savaglio05} used the Gemini Deep Deep Survey (GDDS)
to examine the M-Z relationship at $0.4 < z < 1.0$ and found clear
evidence for an overall decrease in the normalization of the
relationship with respect to that found in the local Universe.
Likewise, \citet{Erb06} utilized 87 rest-frame UV selected star
forming galaxies to study the nature of the correlation beyond a
redshift of 2 and came to similar conclusions.

For our analysis, we have adopted the empirical model put forth by
\citet{Savaglio05} to describe the evolution of the M-Z relationship
as a function of redshift.  This model was developed using their
$0.4 < z < 1.0$ GDDS sample along with the $z \sim 2$ galaxies
presented by \citet{Shapley04} to extrapolate the shape of the M-Z
relationship to higher redshifts.  This empirical relationship \citep[Equation 11 in][]{Savaglio05} allows
for the average metallicity of a galaxy to be estimated as a function
of stellar mass at a given redshift and can be stated as:
\begin{eqnarray} \label{eq:MZRz}
12+{\rm log (O/H)_{\rm KK04}} = -7.59+2.53~{\rm log}~M_{\star}~~~~~~~~~~~~~~ &  \nonumber \\
-0.097~{\rm log}~M_{\star} +5.17~{\rm log}~t_{\rm H}~~~~~~~~~~ &  \nonumber \\ 
 - 0.39~{\rm log}~t_{\rm H} -0.40~{\rm log}~t_{\rm H}~{\rm log}~M_{\star},~~~ \\
 \nonumber
\end{eqnarray}
where $t_{\rm H}$ is the Hubble time and $M_{\star}$ is the galactic
stellar mass.  The Savaglio et al$.$ model reproduces several of the
predicted M-Z relationship properties at high redshift, including the
overall reduction in the M-Z relationship normalization as well as a
steeper evolution in the metallicity of low mass galaxies in
comparison to their high mass counterparts.  Figure 1 shows the
metallicity as a function of stellar mass for a variety of redshifts
as approximated by the Savaglio et al$.$ model out to $z = 5$.  We note that the original data used by \citet{Savaglio05} was limited to the range of $8.2 <$ 12+log(O/H)$_{\rm KK04} < 9.1$ and $0.4 < z \lesssim 2.0$ and hence the curves at lower metallicities and higher redshifts are extrapolations beyond the range of the data used to define the model.   

\begin{figure}
\includegraphics[width=\columnwidth, angle=0]{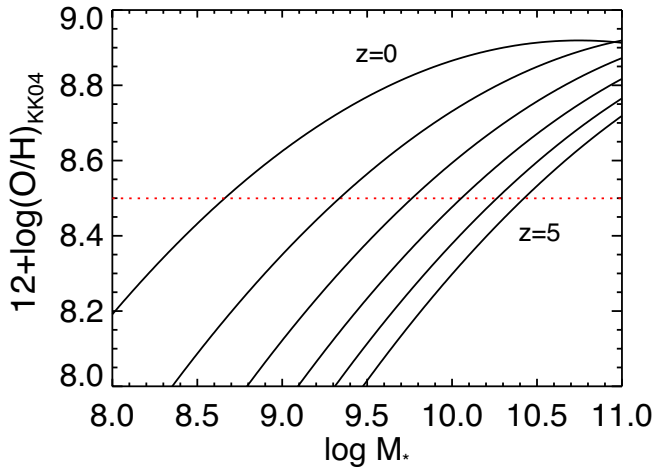}
\caption{The evolution of the galaxy mass-metallicity relationship described by \citep{Savaglio05}, extrapolated to redshifts between $0 < z < 5$.  The overall normalization of the relationship is expected to fall with redshift as metal abundance become less common in all galaxies.  Differential enrichment between low and high mass galaxies also leads to an evolution of the relationship's slope.  The red dotted line represents a low metallicity cutoff of 12+log(O/H)$_{\rm KK04} = 8.5$.  Note that our use of Equation 1 beyond $z ~ 2$ is an extrapolation that is beyond the range of the original data used to define the model.
\label{Fig:MZR}}
\end{figure}

It is important to examine the details of the diagnostics used by
\citet{Savaglio05}, as different initial
mass functions (IMFs), for example, can yield factor of 2 differences in stellar
mass and different metallicity calibrators can likewise result in large
discrepancies in abundance estimates.  The stellar masses used by
\citet{Savaglio05} to produce their empirical relationship were
estimated through SED modeling of multi-band photometry for each
galaxy, with an initial mass function derived by \citet{Baldry03}.  Their metallicity values were obtained through nebular oxygen
abundance estimates calibrated via stellar population synthesis and photoionization
models developed by \citet[hereafter KK04]{KK04}.  This metallicity diagnostic, which uses traditional strong emission line ratios, and
other commonly used calibrations (e.g., \citealt{McGaugh91},
 \citealt{KD02}) are discussed in detail in \citet{Kewley08} who
quantify the systematic offsets amongst the different calibrations and
provide conversion tables.

We also note that the metallicity value for the boundary between hosts that harbor broad-lined SN Ic with associated GRBs and SN Ic with no detected gamma-ray activity reported by \citet{Modjaz08} was measured using the diagnostic proposed by \citet{KD02} (KD02).  In order to convert from the KD02 scale to the KK04 scale used by \citet{Savaglio05}, we consulted \citet{Kewley08} for the appropriate metallicity calibration conversion (their table 3).  We find that a value of 12+log(O/H)$_{\rm KD02} = 8.5$ approximately converts to 12+log(O/H)$_{\rm KK04} \sim 8.66$, which we quote as the \citet{Modjaz08} cutoff metallicity for the remainder of the paper.

In addition to understanding how the average metallicity of a galaxy
varies as a function of stellar mass, we would also like to know how
the number density of galaxies and the number of stars being produced
in those galaxies varies with galactic stellar mass.  This will allow
us to model the effects of a metallicity bias on the overall mass
distribution of GRB host galaxies, and eventually compare those models
to the unbiased mass distribution of all star-forming galaxies at a
given redshift.  As with the mass-metallicity relation, both the
galactic stellar mass function and the star formation rate as a
function of stellar mass are expected to evolve with redshift and
quantifying this evolution is crucial to understanding the
distribution of galaxies that are capable of harboring a GRB.


\begin{figure}
\includegraphics[width=\columnwidth, angle=0]{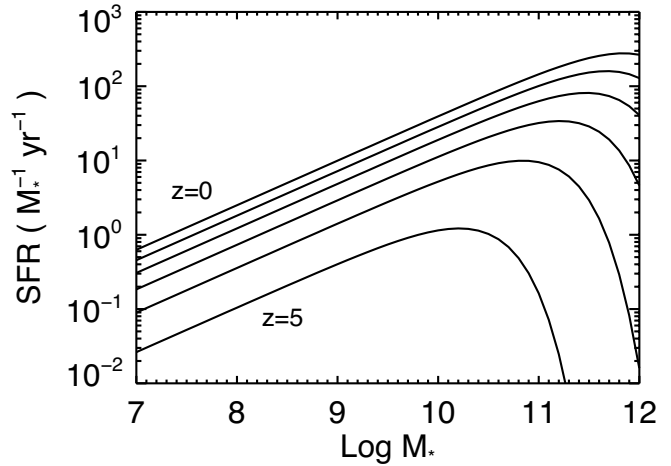}
\caption{The star formation rate as a function of stellar mass between $z = 0 - 5$ as described by \citet{Drory08}.  The stellar mass at which this rate turns over evolves smoothly to higher masses with increasing redshift.}
\label{Fig:SFRvM}
\end{figure}

The star formation rate as a function of stellar mass (SFRM) in the
local universe is well constrained.  Using a sample of more than
$10^{5}$ galaxies, \citet{Kauffmann04} showed that the star formation
rate in low mass galaxies scales as a power law to their halo mass,
peaking at roughly log $M_{\star} \sim 10.4~M_{\odot}$, before
falling for higher mass galaxies.  This transition
represents the stellar mass at which the galaxy distribution changes
from younger stellar populations and active star forming galaxies to
systems with older stellar populations and low star formation
activity.

\citet{Drory08} used the FORS Deep Field survey \citep{Feulner05} to
quantify this relationship and its evolution with redshift for stellar
masses and redshifts spanning $9 <$ log $M_{\star} < 12$ and $0 < z <
5$.  They find that the stellar mass at which the star formation rate
turns over for high mass galaxies evolves smoothly to higher masses
with increasing redshift, until the break mass disappears entirely and
the star formation rate as a function of stellar mass can be
represented as a single power law.  Surprisingly, \citet{Drory08} find that the
power law index representing the low mass region of this relationship
remains constant even to the highest redshifts in their sample.

For the purposes of this paper, we have adopted the analytic
expression presented by \citet{Drory08} for the star formation
rate as a function of stellar mass given as:
\begin{equation}
\dot{M_{\star}}(M_{\star}) = \dot{M}_{\star}^0 \left(\frac{M_{\star}}{M_{\star}^{1}}\right)^{\beta}{\rm exp}\left(-\frac{M_{\star}}{M_{\star}^{1}}\right),
\end{equation}
where M$_{\star}^{1}$ represents the break mass at which the star
formation rate deviates from a power law.  We also use the best fit
parameterizations from Drory et al. (2008) for the evolution of the
overall normalization and break mass with redshift, given as:
\begin{eqnarray} \label{eq:SFRz}
\dot{M_{\star}^{0}} \approx  3.01(1+z)^{3.03}~~~~~~ &  \\
M_{\star}^{1} \approx 2.7\times10^{10}(1+z)^{2.1}
\end{eqnarray}
Following \citet{Drory08}, we have fixed the power law index to $\beta = 0.6$ and assume it
remains constant at all redshifts under consideration.  The star
formation rate as a function of stellar mass between $0 < z < 5$, as
described by Equations (2)-(4), are shown in Figure \ref{Fig:SFRvM}.

\begin{figure}
\includegraphics[width=\columnwidth, angle=0]{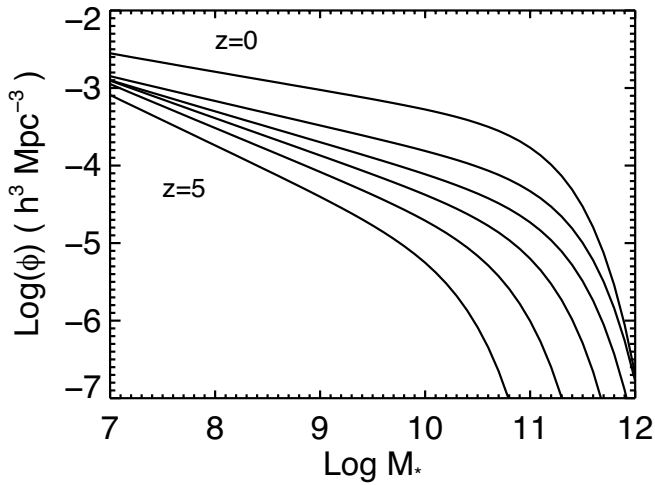}
\caption{The galactic mass function as a function of stellar mass
  between $z = 0 - 5$.  The number density of galaxies decrease as a
  power law in stellar mass before falling sharply at a characteristic
  mass.  The overall number density of high mass galaxies drops
  significantly with redshift.} 
\label{Fig:PhivM}
\end{figure}


The galactic stellar mass function (GSMF) in the local universe is
likewise well understood. It has long been known that dwarf galaxies
represent the largest fraction of galaxies in the local universe, with
their relative number decreasing as a power law with increasing
stellar mass up to some characteristic mass, above which the number of
galaxies drops sharply.  At low redshift, the 2dF \citep{Cole01} and
2MASS-SDSS \citep{Bell03, Blanton03} surveys constrained the
parameters of the Schechter function that is commonly used to describe
the distribution of stellar mass in the
Universe. 
The GSMF at high redshift has been explored by \citet{Fontana04},
\citet{Drory05}, \citet{Conselice05}, and \citet{Fontana06} using a
variety of deep surveys, all showing evidence for a distinct evolution
of the GSMF with cosmic time.  Using the GOODS-MUSIC catalog of over
3000 infrared selected galaxies, \citet{Fontana06} showed that the
number density of high mass galaxies drops with redshift, while the
density of low mass galaxies evolves faster than their high mass
counterparts out to a redshift of $z \sim 1.5$.  The net result of
this differential evolution is an increasing fraction of low mass
dwarf galaxies with respect to higher mass galaxies at higher
redshifts.

\begin{figure}
\includegraphics[width=\columnwidth, angle=0]{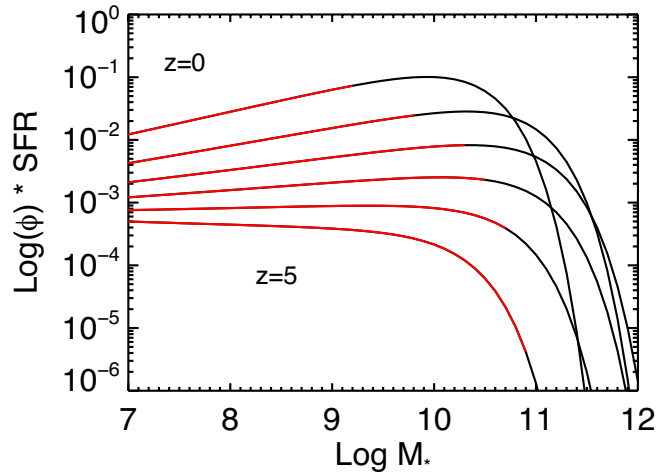}
\caption{The total star formation rate as a function of stellar mass between $0 > z > 5$.  The portion of the curves highlighted in red represent the stellar mass range below the mass limit imposed by a metallicity cutoff of 12+log(O/H)$_{\rm KK04}=8.5$. 
 }
 \label{Fig:GSFM*SFRM}
\end{figure}

For the purposes of this paper, we have adopted the analytic model
presented by \citet{Drory08} for the galactic stellar mass
function given as:
\begin{equation}
\phi(M){\rm d}M = \phi^{\star} \left(\frac{M}{M^{\star}}\right)^{\alpha}{\rm exp}\left(-\frac{M}{M^{\star}}\right)\frac{{\rm d}M}{M^{\star}}.
\end{equation}
We have again used their best fit parameterizations for the evolution
of normalization of the mass function as well as the characteristic
break mass, given as:
\begin{eqnarray} \label{eq:Mz}
\phi^{\star}(z) \approx 0.0031 (1+z)^{-1.07}~~~~ &  \\
{\rm log} M^{\star}(z) \approx 11.35 - 0.22~{\rm ln}(1+z)
\end{eqnarray}
We further assumed that the power law index below the break mass
remains constant at $\alpha = -1.3$ for all redshifts under
consideration.  The GSMF between $0 < z < 5$, as described by
Equations (5)-(7), are shown in Figure \ref{Fig:PhivM}.

Ultimately, it is important to know the total number of stars being
produced as a function of stellar mass. Thus, we also computed the product
of the galactic stellar mass function (GSMF) and star formation rate
as a function of stellar mass (SFRM).
This galaxy weighted star formation rate (WSFR) is shown in Figure
\ref{Fig:GSFM*SFRM} at a variety of redshifts.  The red lines in
Figure \ref{Fig:GSFM*SFRM} represent the metallicity biased WSFR, the
details of which we will discuss in the next section.  Between roughly $0 < z < 3$, the number density of low mass galaxies outweighs that of
their more massive counterparts, but the cosmic star formation rate is
largely dominated by these relatively less numerous massive galaxies.
The net result is a weighted star formation rate that peaks at
intermediate masses, roughly between $10^{10} - 10^{11} M_{\odot}$.
At higher redshifts, the drop in the number of high mass galaxies
becomes significant and the stellar mass function becomes dominated by
low mass galaxies.  At the same time the peak in the SFRM decreases
smoothly to lower masses with increasing redshift, resulting in a
sharp fall in the mass at which the weighted star formation rate peaks for $z > 3$.  The mass at which the WSFR peaks is
plotted as the long dashed black line in Figure \ref{Fig:HostMasses}.
If GRBs are unbiased tracers of star formation in the universe, and if they follow the M-Z relationship \citep[but see][]{Brown08}, then
the peak of their host mass distribution should roughly follow this
line.  We test this prediction in the following section.



\section{Results} \label{sec:Results}

\begin{figure}
\includegraphics[width=\columnwidth, angle=0]{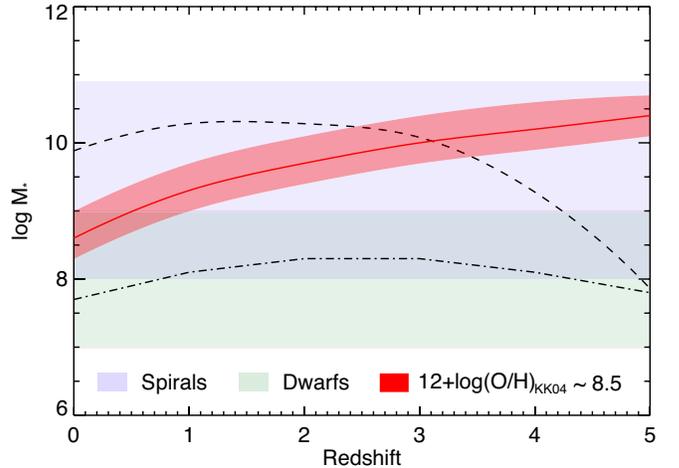}
\caption{The upper limit on the stellar mass of a GRB host galaxy given a sharp metallicity cut-off of 12+log(O/H)$_{\rm KK04} \sim 8.5$ with respect to the typical stellar mass ranges for spiral and dwarf galaxies in the local universe.  The light red region represents the scatter in this limit imposed by the $1\sigma$ scatter of the M-Z relationship at low redshift.  The dashed line represents the stellar mass at which the total star formation in the universe peaks at a given redshift.  Alternatively, the dash-dotted line represents the median stellar mass of this distribution, truncated by the upper limit set by a metallicity bias.   The two regions representing the spiral and dwarf galaxy masses overlap between $10^{9} - 10^{11} M_{\odot}$.}
\label{Fig:HostMasses}
\end{figure}

Using the empirical M-Z relationship expressed in Equation 1, we
estimated the stellar mass of a galaxy of a given
metallicity as a function of look back time.  The average stellar mass
for galaxies with a low oxygen abundance of roughly 1/3 $Z_{\odot}$\footnote[1]{We assume a solar metallicity of 12+log(O/H) = 8.7 \citep{Asplund05} in the \citet{Pettini04} scale and convert that value to the KK04 scale using Table 3 in \citet{Kewley08} to get $Z_{\odot} = 9.0$.}, or 12+log(O/H)$_{\rm KK04} = 8.5$,  is traced as the red line in Figure \ref{Fig:HostMasses}, with the red shaded region surrounding this line representing the
uncertainty due to the intrinsic scatter of the M-Z relationship at
low redshift.  Here we have used the values presented by Tremonti et
al. (2004) to estimate the $1\sigma$ scatter about the M-Z
relationship, and hence the resulting stellar mass range, at a
redshift of $z \sim 0.1$.  Unfortunately, such detailed estimates of
the scatter associated with the M-Z relationship at high redshift are
currently lacking and therefore for our analysis we assume that this
scatter is indicative of the scatter at all redshifts under
consideration, which is certainly an over-simplification.  The region
of stellar mass shaded blue and green represent the typical masses for
dwarf and spiral galaxies respectively.  As expected, the average mass
of a galaxy at a given metallicity rises as a function of redshift, a
direct effect of the decreasing normalization of the M-Z relationship
as a function of look back time.  The effects this would have on a
metallicity biased GRB host population are immediately apparent. If
GRBs are limited to low metallicity environments, then at low redshift
they would be relegated to dwarf and low mass spiral galaxies (barring the effects of metallicity gradients, which we will discuss in detail in $\S 5$"), whereas at high redshift the effective mass limit is raised, allowing GRBs to
occur in a much broader range of galaxies.  A similar prediction was made by \citet{Enrico02} based purely on theoretical grounds. 

Furthermore, if GRBs are unbiased tracers of star formation throughout
the universe, then their observed host mass distribution should
cluster about the peak in the WSFR represented by the long dashed line
in Figure \ref{Fig:HostMasses}.  On the other hand, if they are
metallicity biased tracers of star formation, then their host mass
distribution should deviate significantly from this curve at low
redshifts, peaking instead at the upper mass limit shown in red.  At
some redshift (roughly $z \sim 3$ for a metallicity cutoff of 12+log(O/H)$_{\rm KK04} = 8.5$) the stellar mass at which the WSFR peaks crosses
this upper mass limit, after which the peak in the unbiased and biased
mass distributions become indistinguishable.

We also quantified the median of the GRB host mass distribution by
considering only the product of the GSMF and SFRM below the
metallicity imposed upper mass limit to the GRB host population.  The
red solid lines shown in Figure \ref{Fig:GSFM*SFRM} represent the
galaxy masses which fall below the upper limit imposed by a
metallicity cut-off of 12+log(O/H)$_{\rm KK02}=8.5$ for various
redshifts.  If GRBs are metallicity biased tracers of the star
formation in the universe, then the centroid of this truncated total
star formation rate would
yield the expected median stellar mass of a GRB host population as a
function of redshift.  We plot this expected median mass as a
dash-dotted black line in Figure
\ref{Fig:HostMasses}. 
Although the upper limit imposed on the mass of a galaxy that can host
a GRB increases with redshift, the effects of a galaxy population
dominated by low mass galaxies along with the shift in the type of
galaxies producing most of the stars in the early universe have the
net effect of keeping the median GRB host galaxy mass relatively
constant with redshift, at roughly a mass of $10^{8}~M_{\odot}$.  Note
that this estimate assumes that the fraction of the total star
formation that goes into the production of GRB progenitors does not
change significantly with redshift, host type, or stellar mass.  This
assumption breaks down if environmental variables other than
metallicity play an important role in the formation of a GRB
progenitor.  
normalization in the GRB host mass distribution, only their relative
distribution in stellar galactic mass at a given redshift.

Unfortunately, the predicted median host mass shown by the dash-dotted line if Figure \ref{Fig:HostMasses} is not currently observable, except for low redshift GRBs, which are
rare.  Detection effects and Malquist type biases will lead any
observational measure of the GRB host mass distribution to be biased
towards high mass, high surface brightness, galaxies.  This would
effectively shift the dash-dotted line in Figure \ref{Fig:HostMasses}
to higher masses with increasing redshift and such completeness
considerations have not been incorporated into our
model.  

\section{Comparison to GRB Host Galaxy Observations}

How do the upper mass limits as inferred from the M-Z relationship
compare to measured values for the subset of the GRBs with known host
associations?  To examine this question, we turned to two recent studies
by \citet[hereafter CC08]{Ceron08} and \citet[hereafter
SGB09]{Savaglio09}, which compiled the galactic stellar masses, star
formation rates, and dust extinctions for a large sample of GRB host
galaxies between $0 < z < 2$.  CC08 utilized the rest frame $K$-band
flux densities as interpolated from {\em Spitzer's} \citep{Werner04}
IRAC \citep{Fazio04} and additional NIR observations to obtain an
estimate of $M_{\star}$ for a sample of 30 long-duration GRBs.  SGB09
obtained similar estimates through a combination of optical and NIR
observations for a sample of 46 host galaxies.  Both
groups used photometric observations in conjunction with mass-to-light ratios derived from SED fits to measure the total stellar mass of the hosts in their sample.  The two studies assumed slightly different initial mass
functions (IMFs) and average mass-to-light ratios, introducing a
systematic offset between the estimated mass values derived from the
two samples which we discuss in more detail in the next section.

\begin{figure}
\includegraphics[width=\columnwidth, angle=0]{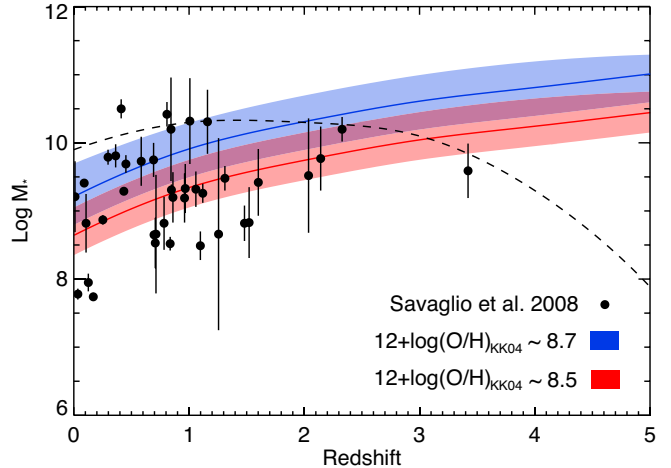}
\caption{Upper limits on the stellar mass of a GRB host galaxy given a metallicity cut-off of 12+log(O/H)$_{\rm KK04} = 8.5$ (red line) and 8.7 (blue line) compared to the masses of 46 host galaxies estimated by \citet{Savaglio09}.  The dashed line represents the stellar mass at which the total star formation in the universe peaks at a given redshift.}
\label{Fig:HostMasses-Savaglio}
\end{figure}

\begin{figure}
\includegraphics[width=\columnwidth, angle=0]{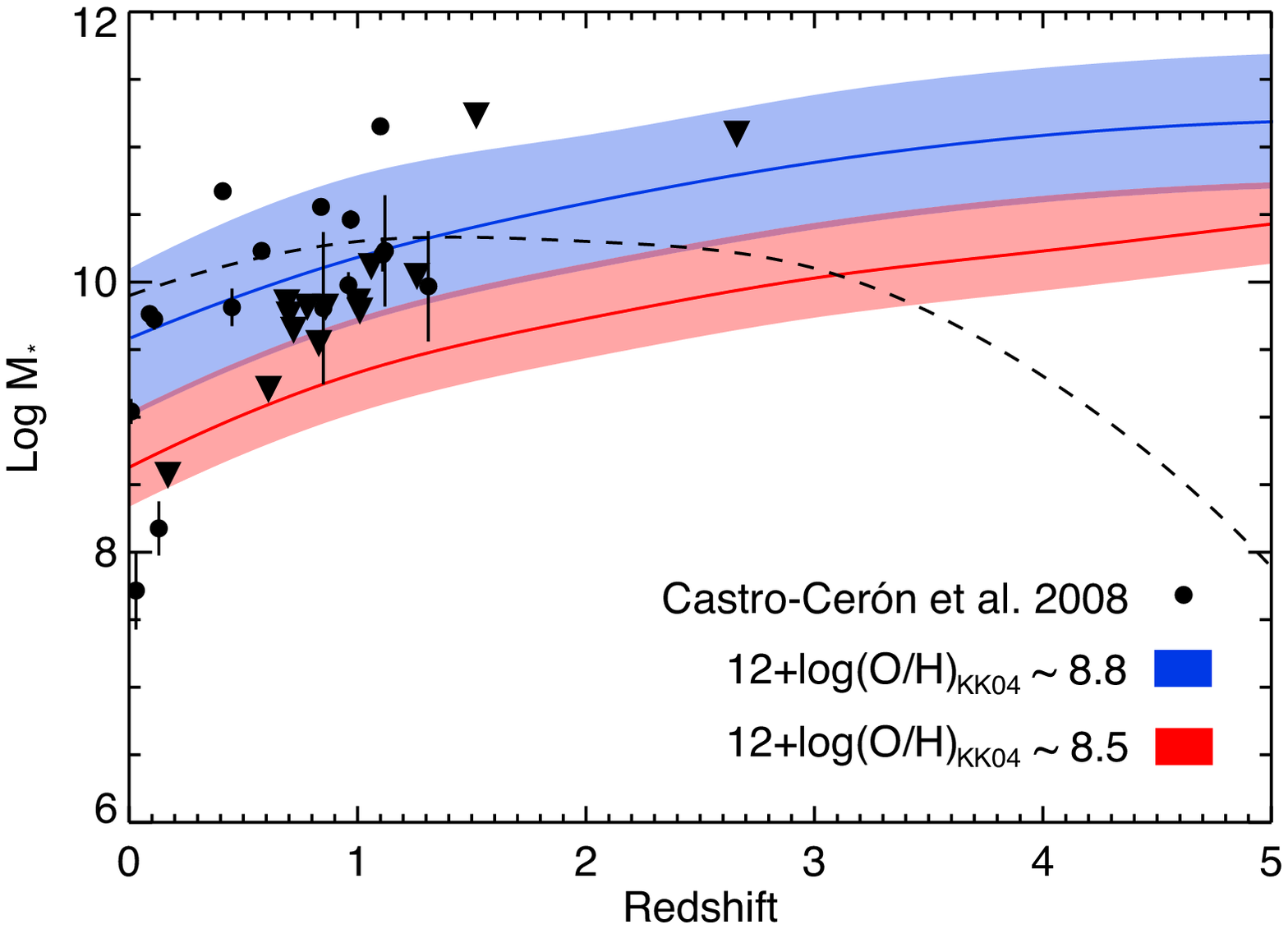}
\caption{Upper limits on the stellar mass of a GRB host galaxy given a metallicity cut-off of 12+log(O/H)$_{\rm KK04} = 8.5$ (red line) and 8.8 (blue line) compared to the masses of 30 host galaxies estimated by \citet{Ceron08}.  The dashed line represents the stellar mass at which the total star formation in the universe peaks at a given redshift.}
\label{Fig:HostMasses-Ceron}
\end{figure}

CC08 and SGB09 found that GRB host galaxies exhibit a wide range of
stellar mass and star formation rates, although as a whole they tend
towards low $M_{\star}$, relatively dim, high specific star-forming
systems, confirming previous observations \citep{Fruchter99,
  LeFloch03, Chary02, Berger03}.  The $M_{\star}$ values from the CC08
and SGB09 papers are shown in Figures \ref{Fig:HostMasses-Savaglio} and
\ref{Fig:HostMasses-Ceron} respectively, with upper limits represented
by triangular symbols.  As in Figure \ref{Fig:HostMasses}, the red
shaded region in both plots represents the upper limit on the stellar
mass of a galaxy capable of hosting a GRB as imposed by the M-Z relationship and its
associated $1\sigma$ scatter with a metallicity cut-off of 12+log(O/H)$_{\rm KK04} = 8.5$   It is clear from Figures
\ref{Fig:HostMasses-Savaglio} and \ref{Fig:HostMasses-Ceron} that a
significant fraction of observed host galaxies have $M_{\star}$ values
that are greater than the predicted upper limit to the GRB host mass
distribution for such a low metallicity cutoff value.  Most of these
host galaxies can be accommodated if the metallicity cut-off is
increased to 12+log(O/H)$_{\rm KK04} = 8.7$ for the SGB09 sample and 12+log(O/H)$_{KK04} = 8.8$ for the CC08 sample.  Note that the resulting spread
in the predicted mass range is significantly wider for 12+log(O/H)$_{\rm KK04}
= 8.8$ due to the shallower slope of the shallower M-Z relationship at
this metallicity.  Even at this relatively high metallicity cut-off,
with its larger intrinsic spread, three hosts in the CC08 sample are
still above the predicted mass limit, although metallicity gradients within these high-mass hosts may explain the existence of these hosts in the GRB sample.

SGB09 and CC08 found median masses of $M_{\star} \sim 10^{9.3}
M_{\odot}$ and $M_{\star} \sim 10^{9.7} M_{\odot}$ respectively, far
greater than the median host mass predicted by simply looking at the
truncated distribution of total star formation as a function of
$M_{\star}$ (The dash-dotted line in Figure \ref{Fig:HostMasses}).
This direct comparison between the expectation peak of the WSFR and
the median values for the two samples is problematic, as detection
effects biasing against low mass galaxies will heavily influence the
observed median mass.  We can however compare the observed host mass
distribution to the high end of the expected mass distribution of all
star formating galaxies at a given redshift, as detection effects
should not effect this region of the observed distribution.  We
address this question in Figures \ref{Fig:Host_Dists_Savaglio} and
\ref{Fig:Host_Dists_Ceron}, where we plot the SGB09 and CC08 host mass
distributions for galaxies between $0.75 < z < 1.25$ along with the
expected unbiased WSFR as a function stellar mass at a $z = 1$ (dashed line).  The normalization of the distributions in these
plots is arbitrary, with the peak of the predicted WSFR and
the observed host mass distributions both being set to 1.  The
bracketed arrows in Figure \ref{Fig:Host_Dists_Ceron} represent
galaxies for which CC08 were unable to make firm estimates on the
galactic stellar mass, resulting only in upper limits accompanied with
very conservative lower limits.


It is quite clear that the SGB09 sample is not well described by the
expected host mass distribution of unbiased star forming galaxies at $z =
1$.  Although the SGB09 distribution can be expected to artificially
fall off at low $M_{\star}$ due to observational biases, the same
cannot be said for the lack of high $M_{\star}$ galaxies, pointing to
an intrinsic decline in the GRB host population.  The case for the
CC08 sample is less clear.  Their median stellar mass between $0.75 <
z < 1.25$ of $M_{\star} = 10^{10.23} M_{\odot}$ is much more
consistent with the peak of the unbiased WSFR distribution, which at
$z \sim 1$ peaks at $M_{\star} = 10^{10.30} M_{\odot}$.  This median
of the CC08 sample does not include the values for which only limits
exist, which work to broaden the distribution to lower $M_{\star}$
values, making it less consistent with the model distribution.

We can statistically compare the two observed distributions to the
model distribution by drawing a random set of values from the WSFR
distribution, equal in size to the observed samples, to which we can
perform a two-sided Kolmogorov-Smirnov analysis.  We perform this
comparison for 1000 trails, using a random realization of the WSFR
mass distribution in each iteration, while measuring the median
probability that the model distribution and the SGB09 and CC08 samples
are drawn from the same parent populations.  For both the SGB09 and
CC08 samples, the probability that they are randomly drawn from the
unbiased WSFR distribution is quite low, $6.3\times10^{-12}$ and
$1.6\times10^{-05}$ respectively.  Unfortunately, the observational
biases discussed above leads to the lack of completeness at low
$M_{\star}$ for both samples making the use of a traditional K-S
analysis questionable.  The median WSFR mass will be heavily weighted
by low mass galaxies, the smallest of which are likely not present in the SGB09
and CC08 samples because of detection effects. 

At present, without an understanding of the completeness of the GRB
host samples, we can only compare the peaks and the high end behavior
of the mass distributions which we believe should not be effected by
observational biases.  In both cases, the SGB09 and CC08 samples peak
below the unbaised peak in the galaxy weighted star formation rate as
a function of stellar mass, although the discrepancy is much greater
when considering the SGB09 sample.

\section{Discussion} \label{sec:Discussion}

The comparison between the stellar mass limits imposed by metallicity
cut-offs to the measured $M_{\star}$ values in the SGB09 and CC08
samples is quite telling.  A low metallicity cut-off of 12+log(O/H)$_{\rm KK04} = 8.5$ is disfavored by current measurements of the stellar
masses of GRB host galaxies at low and intermediate redshifts.
However, a comparison of observed GRB host masses still appears to favor a metallicity biased mass
distribution rather than one based solely on the mass distribution of
all star formation galaxies at similar redshifts. Increasing the
metallicity cut-off to 12+log(O/H)$_{\rm KK04} \sim 8.7-8.8$ allows
for the accommodation of most of measured host masses, when factoring
in the intrinsic spread of the M-Z relationship.  This is in rough agreement with the metallicity cutoff found by \citet{Modjaz08} of roughly 12+log(O/H)$_{\rm KK04} \sim 8.66$ at low redshift ($z < 0.25$).  This result is also in general agreement with the results presented by \citet{Nuza07} who
find a comparable metallicity bias through the use of hydrodynamical
cosmological simulations in conjunction with assumptions of the
collapsar event rate.  They conclude that the observed properties of
GRB host galaxies are reproduced if long GRBs are limited to
low-metallicity progenitors.

In a similar investigation, \citet{Wolf07} used the
luminosity-metallicity ($L-Z$) relation for galaxies to compare the
host galaxy luminosity distributions between CC SNe and long GRB host
galaxies to the expected luminosity function of all star forming
galaxies at a given redshift.  They found that although their
ultraviolet based SFR estimates reproduced the CC SNe host luminosity
distribution extremely well, the same was not true for the GRB host
population.  They found that their model SFR estimates would have to
exclude luminous, and hence high metallicity, galaxies in order to
match the observed GRB host distribution.  They concluded that a
metallicity bias with a cutoff of roughly 12+log(O/H)$_{\rm KK04} \sim 8.7$
would be sufficient to reproduce the observed distribution, although
they stressed that they could not distinguish between a sharp cutoff or a
decreasing efficiency at producing GRBs as a function of decreasing
host metallicity with their current data.  This decreasing efficiency
is more realistic than a sharp efficiency cutoff and, combined with
the spread in the M-Z relationship, could explain the existence of
outliers in Figures \ref{Fig:HostMasses-Savaglio} and
\ref{Fig:HostMasses-Ceron}.

Metallicity gradients within galaxies also work to dilute observable
evidence for a sharp metallicity cut-off in the galaxies that can
harbor GRBs.  The metallicities within disk galaxies tend to fall as a
function of radius from the core \citep[e.g.][and references therein]{Kewley05} and as such the host
integrated light represents an upper limit to the metallicity at the
GRB location.  The nearby galaxies that most closely resemble a
typical GRB host galaxy at low $z$ for which we have spatially
resolved spectroscopy are the Large and Small Magellanic Clouds, both
of which have small internal dispersions of order 0.1 dex in oxygen abundance
\citep{Russell92}.  This value is common for star forming dwarf irregular galaxies in which metallicity gradients are rather negligible, although the internal dispersion as measured from HII regions in larger galaxies such as the Milky Way can be as high as 0.3 dex \citep{Carigi05, Esteban05}. 

 \begin{figure}
\includegraphics[width=\columnwidth, angle=0]{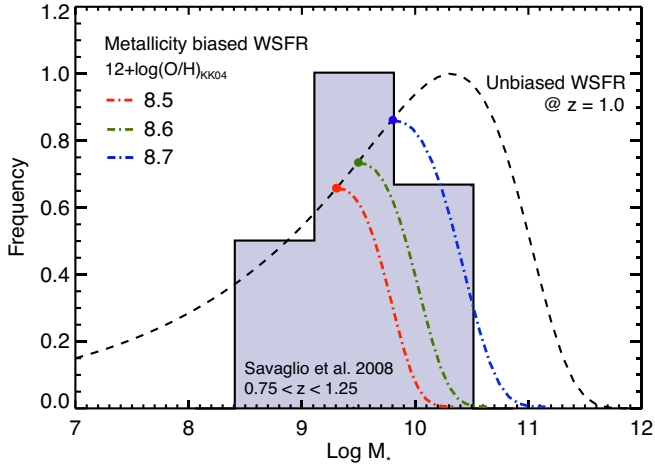}
\caption{The GRB host mass distribution as measured by SGB09 between $0.75 \leqslant z \leqslant 1.25$ compared to the total galaxy weighted SFR as a function of galactic stellar mass at $z \sim 1$ (dashed line).  The mass limits due to sharp metallicity cut-offs of 12+log(O/H)$_{\rm KK04}$ values of 12+log(O/H)$_{\rm KK04}$ = 8.5, 8.6, and 8.7 are represented by a red, green, and blue filled circles respectively.  The combined effects of a smooth efficiency cutoff in the production of a GRB as a function of metallicity are shown as dash-dotted lines proceeding each upper mass limit.  The peak of the SGB09 sample is roughly an order of magnitude below the expected peak of an unbiased galaxy weighted SFR.}
\label{Fig:Host_Dists_Savaglio}
\end{figure}

\begin{figure}
\includegraphics[width=\columnwidth, angle=0]{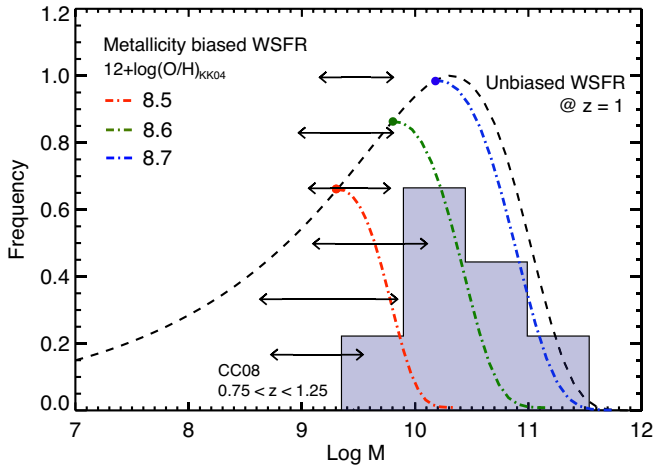}
\caption{The GRB host mass distribution as measured by CC08 between $0.75 \leqslant z \leqslant 1.25$ compared to the total galaxy weighted SFR as a function of galactic stellar mass at $z \sim 1$ (dashed line).  The mass limits due to sharp metallicity cut-offs of 12+log(O/H)$_{\rm KK04}$ values of 12+log(O/H)$_{\rm KK04}$ = 8.5, 8.7, and 8.8 are represented by a red, green, and blue filled circles respectively.  The arrows represent galaxies for which CC08 could only estimate conservative upper and lower limits to their mass.   The combined effects of a smooth efficiency cutoff in the production of a GRB as a function of metallicity are shown as dash-dotted lines proceeding each upper mass limit.  The CC08 mass distribution is much broader than the SGB09 sample at this redshift, although the galaxies for which only upper and lower limits exists pull the peak of their distribution between $10^{9}-10^{10} M_{\odot}$, below the expected peak of an unbiased galaxy weighted SFR.}
\label{Fig:Host_Dists_Ceron}
\end{figure}

The combined effects of a smooth efficiency cutoff and this relatively
small expected metallicity gradient on the GRB mass distribution are
shown as dash-dotted lines in Figures \ref{Fig:Host_Dists_Savaglio}
and \ref{Fig:Host_Dists_Ceron}.  The upper mass limits due to sharp
metallicity cut-offs are marked by the filled red, green, and blue
dots as labeled.  The dash-dotted lines proceeding each limit
represents the unbiased WSFR distribution convolved with a smoothly
broken power law decline in the efficiency of producing a GRB at a
given metallicity, and hence stellar mass.  Any effect of a
metallicity gradient in a typical host galaxy would work to extend
the peak of the Z-biased mass distribution to higher masses.  We assumed that at low
$M_{\star}$ values, the cutoff efficiency is 1, transitioning sharply as
$M_{\star} \rightarrow M_{\rm cutoff}$ to a power law decline of index
$\alpha = -1$.  We believe that such a power-law index can accommodate
the spread in allowable metallicities from both the effects of a
declining efficiency and the small metallicity dispersion expected in GRB host galaxies.


In the context of these two effects, the resulting $M_{\rm cutoff}$
now can be understood as the peak in the predicted GRB host mass
distribution at low redshift and not a sharp upper limit. As such,
this smooth decrease in efficiency can accommodate host galaxies of
much higher stellar mass than the scatter in the M-Z relationship
alone.  At a metallicity cut-off of 12+log(O/H)$_{\rm KK04} = 8.7$, for
example, galaxies of $M_{\star} \sim 10^{11} M_{\odot}$ are permitted
by the model (in relative abundance) whereas the scatter in the M-Z
relationship with a sharp efficiency cut-off would strictly exclude
galaxies above $M_{\star} \sim 10^{10}$ at a $z \sim 1$.  

A cut-off of 12+log(O/H)$_{\rm KK04} \sim 8.7$, or 1/2 $Z_{\odot}$, does contain most of the CC08
sample, although the low mass location of the peak in the SGB09 sample
points to a systematic difference between the two samples.  The
two host samples have a total of 25 overlapping objects and CC08
discussed the differences between the two studies in some detail.   They concluded that the higher median mass in their
sample reflects a lower mass-to-light ratio obtained from the subset of galaxies for which they performed SED fits, compared to the average value obtained by SGB09 through SED fitting to their entire host sample.  They also added that the use of
optical-NIR SEDs by SGB09 may underestimate the effects of dust
extinction for obscured galaxies, giving rise to further
discrepancies.

We also note that CC08 used a traditional Salpeter IMF
\citep{Salpeter55} to estimate the relative number of low mass, and
hence undetected, stars within a galaxy, whereas SGB09 utilized a
modified Salpeter IMF presented by \citet{Baldry03}.  The host masses
derived through the use of these two IMFs can differ substantially, as
the traditional Salpeter IMF tends to overestimate the number of
low-mass dwarf stars compared to updated models presented by
\citet{Baldry03}, \citet{Kroupa01}, and \citet{Bochanski}.  We
estimated that masses derived through the use of a Baldry $\&$
Glazebrook IMF are systematically lower by roughly $85\%$ compared to
those found through the use of a Salpeter IMF for a given
mass-to-light ratio.  This combined with the lower effective
mass-to-light ratio used by SGB09 may explain the discrepancies between
the two samples.  It is important to note that \citet{Savaglio05}
explicitly used the Baldry $\&$ Glazebrook IMF to obtain the M-Z
relationship parameterization that we use in Equation \ref{eq:MZRz},
therefore their sample makes for a more meaningful comparisons to our
model.

The effects of a metallicity bias in the GRB progenitor population
combined with an evolving M-Z relationship would suggest that
afterglow associations with massive galaxies of $M_{\star} > 10^{11}
M_{\odot}$ should be limited to high redshifts events.  Unfortunately,
very few host galaxies have measured masses above $z > 2$ to test this
directly, despite the median redshift of Swift detected GRBs of $z
\sim 2.75$, highlighting the difficulty in observing many of these
high redshift hosts.  Despite this increase with redshift in the upper limit in the
mass of galaxies capable of harboring a GRB, we find
that an evolving galaxy populations in which dwarf galaxies represent
a larger fraction of the star forming galaxies in the distance
universe results in a median GRB host mass that remains fairly
constant between $0 < z < 3$  Above $z > 3$, we see this upper limit fall sharply, implying that a large fraction of GRBs at high redshift
will still occur in low mass galaxies.

Finally, if the normalization of the M-Z relationship for galaxies decreases as a function of lookback time, then there should be some redshift at which a metallicity biased galaxy populations would become indistinguishable from the star-forming field galaxy population.  For a metallicity cutoff of 12+log(O/H)$_{\rm KK04} = 8.7$, we find that the peak in the stellar mass distributions between these two populations should become equal at a redshift of $z \sim 2$, with the biased and unbiased populations becoming less distinguishable at higher redshifts.  The greatest discrepancy between a metallicity biased host population at that of the population of all star forming field galaxies would occur at low to intermediate redshifts.  This may explain the discrepancy between high redshift host properties reported by \citet{Chen08} and \citet{Fynbo08} and the properties reported by \citet{Wolf07} for hosts at intermediate redshifts. \citet{Chen08} found that the UV luminosity distribution of long GRB hosts is largely consistent with their being drawn from a UV luminosity weighted random galaxy population at similar redshifts.  \citet{Fynbo08} reported on similar agreements when comparing the luminosity and metallicity distributions of GRB hosts to UV-selected star forming galaxies at $z \sim 3$.  This is in stark disagreement with the conclusions reported by \citet{Wolf07} who find that a metallicity truncated field population is required to match the luminosity distribution of GRB host galaxies at redshifts of $0.4 < z < 1.0$.  This dichotomy between the high redshift and low redshift comparisons would be expected, if at some point, the two populations become indistinguishable as the average metallicity of the field galaxies falls with increasing lookback time. 

\section{Conclusions} \label{sec:Discussion}

We find that dearth of massive GRB host galaxies at low and
intermediate redshifts exceeds that expected from the decline in the
predicted number of massive star forming galaxies at similar
redshifts.  We therefore conclude that there is sufficient evidence to
indicate that GRB host galaxies are metallically biased tracers of
star formation at low and intermediate redshifts and suggest that this
bias should disappear at higher redshifts due to the evolving
metallicity content of the early universe.  We find that a galaxy mass function that includes a smooth decrease in the efficiency of producing GRBs in galaxies of metallicity above 12+log(O/H)$_{\rm KK04}$ = 8.7 accommodates a majority of the measured host masses.  This is in rough agreement with the metallicity cutoff found by \citet{Modjaz08} of roughly 12+log(O/H)$_{\rm KK04} \sim 8.66$ at low redshift ($z < 0.25$).  Throughout our analysis, the modeling and subsequent metallicity comparisons have been performed in the same, consistent fashion and in the same metallicity calibration scale, in order to avoid any systematic differences between the various metallicity diagnostics used in the literature.
  
For a metallicity cut-off of 12+log(O/H)$_{\rm KK04} \sim 8.7$, the predicted peak in the GRB host mass distribution and the stellar mass at which the weighted star formation
rate peaks become equal at $z \sim 2$, with higher values of 12+log(O/H)$_{\rm cutoff}$ pushing this intersection to lower redshift.  This limits
the redshift range in which the differences between a metallicity
biased GRB host population and that of unbiased star forming galaxies can be
tested through direct luminosity or mass distribution comparisons.
Therefore, comparisons of these distributions at low and intermediate
redshifts will be crucial to further inquires into the nature of the metallicity bias in the GRB host population.

\acknowledgements

D.K. acknowledges financial supported through the NSF Astronomy $\&$ Astrophysics Postdoctoral Fellowships under award AST-0502502 and the Fermi Guest Investigator program.  A.A.W. thanks Adam Burgasser for financial support.  M.M. is supported by a research fellowship through the Miller Institute for Basic Research in Science.  We thank Joshua Bloom, Jason X. Prochaska, Hsiao-Wen Chen, and Daniel Perley for insightful discussion and assistance. This work was supported in part by the U.S. Department of Energy contract to SLAC no. DE-AC3-76SF00515

\bibliography{bibtex.bib}

\end{document}